\documentclass{article}
\usepackage{graphicx}
\begin{document} 

\noindent{\huge{\bf Community Structure in Graphs}} 
\vskip1cm
{\large
\noindent{Santo Fortunato$^a$, Claudio Castellano$^b$}
\vskip0.5cm
\noindent{$^a$ Complex Networks Lagrange Laboratory (CNLL), ISI Foundation, Torino, Italy}\newline
\noindent{$^b$ SMC, INFM-CNR and Dipartimento di Fisica, ``Sapienza''
Universit\`a di Roma, P. le A. Moro 2, 00185 Roma, Italy}\newline
}
\vskip1cm

\noindent{\Large{\bf\centerline{Abstract}}} 
\vskip0.3cm

Graph vertices are often organized into groups that seem to live fairly independently
of the rest of the graph, with which they share but a few edges, whereas the relationships
between group members are stronger, as shown by the large number of mutual connections.
Such groups of vertices, or communities, can be considered as independent compartments 
of a graph. Detecting communities is of great importance in sociology, biology and computer science,
disciplines where systems are often represented as graphs. The task is very hard, though, both conceptually, due to the ambiguity in the definition
of community and in the discrimination of different partitions and
practically, because algorithms must find ``good'' partitions among an
exponentially large number of them.
Other complications are represented by the possible
occurrence of hierarchies, i.e. communities which are nested 
inside larger communities, and by the existence of overlaps between communities, due 
to the presence of nodes belonging to more groups. 
All these aspects are dealt with in some detail and many methods are described,
from traditional approaches used in computer science and sociology to recent techniques developed mostly within statistical physics.

\section{Introduction}

The origin of graph theory dates back to Euler's solution~\cite{euler} of the puzzle of 
K\"onigsberg's bridges in 1736. Since then a lot has been learned about graphs and their mathematical properties~\cite{bollobas}. 
In the 20th century they have also become extremely useful as representation of a wide variety of systems
in different areas. Biological, social, technological, and information networks can be studied as
graphs, and graph analysis has become crucial to understand the features of these systems.
For instance, social network analysis started in the 1930's and has become one of the most important
topics in sociology~\cite{wasserman,scott}. 
In recent times, the computer revolution has provided scholars with a huge amount of data
and computational resources to process and analyse these data. The size of real networks 
one can potentially handle has also grown considerably, reaching millions or even
billions of vertices. The need to deal with such a large number of units has produced
a deep change in the way graphs are approached~\cite{bara02}-\cite{vitorep}. 

\begin{figure}
\begin{center}
\includegraphics[width=6cm]{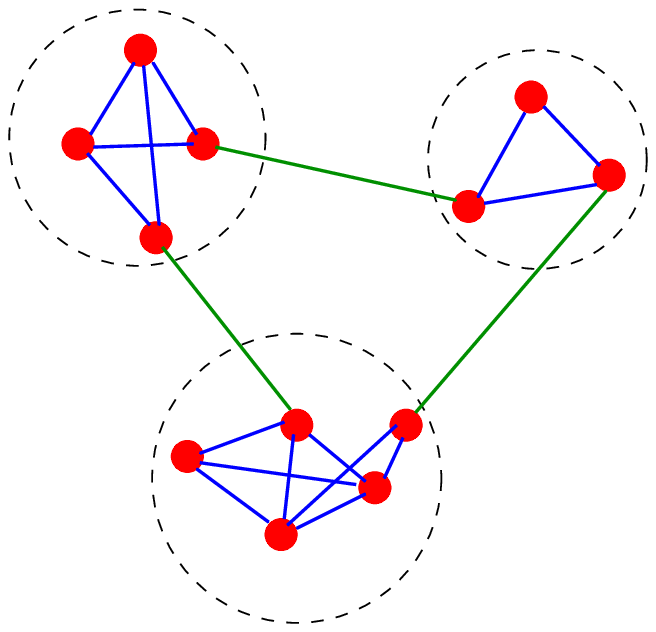}
\caption {\label{Figure1} A simple graph with three communities, highlighted by the
dashed circles.}
\end{center}
\end{figure}

Real networks are not random graphs.
The random graph, introduced by P. Erd\"os and A. R\'enyi~\cite{erdos},
is the paradigm of a disordered graph: in it, the probability 
of having an edge between a pair of vertices is equal for all possible pairs.
In a random graph, the distribution of edges among the vertices is highly homogeneous.
For instance, the  
distribution of the number of neighbours of a vertex, or {\it degree}, is
binomial, so most vertices have equal or similar degree.
In many real networks, instead, there are big inhomogeneities, 
revealing a high level of order and organization. The degree distribution
is broad, with a tail that often follows a power law: therefore, many vertices
with low degree coexist with some vertices with large degree. Furthermore, the 
distribution of edges is not only globally, but also locally inhomogeneous, with 
high concentrations of edges within special groups of nodes, and low concentrations between these groups.
This feature of real networks is called {\it community structure} and is the topic of this chapter.
In Fig.~\ref{Figure1} a schematic example of a graph with community structure
is shown.

Communities are groups of vertices 
which probably share common properties and/or play similar roles within the graph. So,
communities may correspond to 
groups of pages of the World Wide Web dealing with related topics~\cite{Flake:2002}, 
to functional modules such as cycles and pathways 
in metabolic networks~\cite{Guimera:2005, Palla:2005}, to groups of related individuals
in social networks~\cite{Girvan:2002, Lusseau:2003}, to compartments in food webs~\cite{foodw1, foodw2}, and so on.

Community detection is important for other reasons, too. Identifying modules and their boundaries
allows for a classification of vertices, according to their topological position in the modules. So, 
vertices with a central position in their clusters, i.e. sharing a large number of edges with the other
group partners, may have an important function of control and stability within the group; 
vertices lying at the boundaries between modules play an important role of mediation
and lead the relationships and exchanges between different communities. Such classification
seems to be meaningful in social~\cite{granovetter}-\cite{freeman} and metabolic networks~\cite{Guimera:2005}. 
Finally, one can study the graph where vertices are the communities and edges are set between
modules if there are connections between some of their vertices 
in the original graph and/or if the modules overlap. In this way one attains a coarse-grained 
description of the original graph, which unveils the relationships between modules. Recent studies
indicate that networks of communities have a different degree distribution with respect to the
full graphs~\cite{Palla:2005}; however, the origin of their structures can be explained
by the same mechanism~\cite{pollner}. 

The aim of community detection in graphs is to identify the modules only
based on the topology. The problem has a long tradition and it has
appeared in various forms in several disciplines. For instance, in parallel computing it is crucial 
to know what is the best way to allocate tasks to processors so as to minimize the
communications between them and enable a rapid performance of the calculation. 
This can be accomplished by splitting the computer cluster into
groups with roughly the same number of processors, such that the number of physical connections
between processors of different groups is minimal. 
The mathematical formalization of this problem is called
{\it graph partitioning}. The first algorithms for graph partitioning 
were proposed in the early 1970's. Clustering analysis is also 
an important aspect in the study of social networks. The most popular techniques are hierarchical clustering and k-means clustering,
where vertices are joined into groups according to their mutual similarity.

In a seminal paper, 
Girvan and Newman proposed a new algorithm, aiming at the identification
of edges lying between communities and their successive removal, a procedure that after a few iterations
leads to the isolation of modules~\cite{Girvan:2002}. The intercommunity edges are detected 
according to the values of a centrality measure, the edge betweenness, that expresses the importance of the role of 
the edges in processes where signals are transmitted across the graph following paths of minimal length.
The paper triggered a big activity in the field, and many new methods have been proposed in the last years.
In particular, physicists entered the game, bringing in their tools and techniques: spin models, optimization,
percolation, random walks, synchronization, etc., became ingredients of new original algorithms.
Earlier reviews of the topic can be found in Refs.~\cite{newman04rev,arenasrev}. 
 
Section~\ref{sec1} is about the basic elements of community detection, starting from the definition of community. 
The classical problem of graph partitioning and the methods for clustering analysis in sociology are 
presented in Sections~\ref{sec2} and~\ref{sec3}, respectively. Section~\ref{sec4} is devoted to
a description of the new methods. In Section~\ref{sec5} the problem of testing
algorithms is discussed.
Section~\ref{sec6} introduces the description of graphs at the level of communities. Finally, Section~\ref{sec7}
highlights the perspectives of the field and sorts out promising research directions for the future.

This chapter makes use of some basic concepts of graph theory, that can be found in any introductory textbook,
like ~\cite{bollobas}. Some of them are briefly explained in the text.

\section{Elements of Community Detection}
\label{sec1}

The problem of community detection is, at first sight, intuitively clear.
However, when one needs to formalize it in detail things are
not so well defined.
In the intuitive concept some ambiguities are hidden and there are
often many equally legitimate ways of resolving them.
Hence the term ``Community Detection'' actually indicates several
rather different problems.

First of all, there is no unique way of translating into a precise
prescription the intuitive idea of community. Many possibilities
exist, as discussed below. 
Some of these possible definitions allow for vertices to belong to
more than one community. It is then possible to look for overlapping
or nonoverlapping communities.
Another ambiguity has to do with the concept of community structure.
It may be intended as a single partition of the graph
or as a hierarchy of partitions, at different levels of coarse-graining.
There is then a problem of comparison. 
Which one is the best partition (or the best hierarchy)? 
If one could, in principle, analyze all possible partitions of a graph,
one would need a sensible way of measuring their ``quality''
to single out the partitions with the strongest community structure.
It may even occur that one graph has no community structure and one
should be able to realize it.
Finding a good method for comparing partitions is not a trivial task
and different choices are possible.
Last but not least, the number of possible partitions grows faster than exponentially
with the graph size, so that, in practice, it is not possible to
analyze them all. Therefore one has to devise smart methods to find
'good' partitions in a reasonable time. Again, a very hard problem.

Before introducing the basic concepts and discussing the relevant questions
it is important to stress that the identification of topological clusters
is possible only if the graphs are {\it sparse}, i.e. if
the number of edges $m$ is of the order of the number of nodes $n$ of the
graph. If $m\gg n$, the distribution
of edges among the nodes is too homogeneous for communities to make sense.

\subsection{Definition of Community}
\label{seccomm}

The first and foremost problem is how to define precisely what a community is. The intuitive notion
presented in the Introduction is related to the comparison of the number of edges joining 
vertices within a module (``intracommunity edges'') with the number of edges joining vertices of different modules
(``intercommunity edges''). A module is characterized by a larger density of links ``inside'' than ``outside''.
This notion can be however formalized in many ways. Social network analysts have devised many definitions of 
subgroups with various degrees of internal cohesion among vertices~\cite{wasserman,scott}. 
Many other definitions have been introduced by computer scientists and physicists. In general, the definitions can be classified in three main categories. 
\begin{itemize}
\item{{\it Local definitions}. Here the attention is focused on the vertices of the subgraph under 
investigation and on its immediate neighbourhood, disregarding the rest of the graph. 
These prescriptions come mostly from social network analysis and
can be further subdivided in {\it self-referring}, when one considers the subgraph alone, and {\it comparative},
when the mutual cohesion of the vertices of the subgraph is compared with their cohesion with 
the external neighbours. Self-referring definitions identify classes of subgraphs
like {\it cliques}, {\it n-cliques}, {\it k-plexes}, etc.. They are {\it maximal
subgraphs}, which cannot be enlarged with the addition of new vertices and edges without
losing the property which defines them. 
The concept of clique is very important and often recurring when one studies graphs. A clique  
is a maximal subgraph where each vertex is adjacent to all the others. In the literature it is common 
to call cliques also non-maximal subgraphs. Triangles are 
the simplest cliques, and are frequent in real networks. Larger cliques are rare, so 
they are not good models of communities. Besides, finding cliques is computationally very demanding: the 
Bron-Kerbosch method~\cite{bron} runs in a time growing exponentially with the size of the graph.
The definition of clique is very strict. A softer
constraint is represented by the concept of n-clique, which is a maximal subgraph such that the  
distance of each pair of its vertices is not larger than $n$. A k-plex is a maximal subgraph such that each vertex is adjacent
to all the others except at most $k$ of them. In contrast, a {\it k-core} is a maximal subgraph
where each vertex is adjacent to at least $k$ vertices within the subgraph. 
Comparative definitions include that of {\it LS set}, or {\it strong community}, and that of  
{\it weak community}. An LS set is a subgraph where each node has more neighbours inside than outside the
subgraph. Instead, in a weak community, the total degree of the nodes inside the community exceeds the
external total degree, i.e. the number of links lying between the community and the rest of the graph. 
LS sets are also weak communities, but the inverse is not true, in general. The notion of weak community
was introduced by Radicchi et al.~\cite{radicchi}.}
\item{{\it Global definitions}. Communities are structural units of the graph, so it is reasonable to think
that their distinctive features can be recognized if one analyses a subgraph with respect to the graph as a whole. 
Global definitions usually start from a {\it null model}, i.e. a graph which matches the original 
in some of its topological features, but which does not display community structure. 
After that, the linking properties of subgraphs of the initial graph are compared with those of the corresponding 
subgraphs in the null model. 
The simplest way to design a null model is to introduce randomness in the distribution of edges among the vertices.
A random graph \`a la Erd\"os-R\'enyi, for instance, has no community structure, as any two vertices have the same
probability to be adjacent, so there is no preferential linking involving special groups of vertices. The most popular null model
is that proposed by Newman and Girvan and consists of a randomized version of the original graph, where edges are rewired
at random, under the constraint that each vertex keeps its degree~\cite{Newman:2004b}. This null model is the basic concept 
behind the definition of {\it modularity},
a function which evaluates the goodness of partitions of a graph into modules (see Section~\ref{sec01}). 
Here a subset of vertices is a community 
if the number of edges inside the subset exceeds the expected number of internal edges that the subset would have in the null model.
A more general definition, where one counts small connected subgraphs ({\it motifs}), and not necessarily edges,
can be found in ~\cite{arenas07d}.
A general class of null models, including modularity, has been designed by 
Reichardt and Bornholdt~\cite{Reichardt:2006}.} 

\item{{\it Definitions based on vertex similarity}. In this last category, communities are groups of vertices which are 
similar to each other. A quantitative criterion is chosen to evaluate the similarity between each pair of vertices, connected or not. The criterion
may be local or global: for instance one can estimate the distance between a pair of vertices. Similarities can be also extracted from 
eigenvector components of special matrices, which are usually close in value for vertices belonging to the same community. Similarity measures are 
at the basis of the method of hierarchical clustering, to be discussed in Section~\ref{sec3}. The main problem in this case is the need to    
introduce an additional criterion to ``close'' the communities.}
\end{itemize}

It is worth remarking that, in spite of the wide variety of definitions, in many detection algorithms communities are not defined at all, but are 
a byproduct of the procedure. This is the case of the divisive algorithms described in Section~\ref{sec41} and of the 
dynamic algorithms of Section~\ref{sec44}.

\subsection{Evaluating Partitions: Quality Functions}
\label{sec01}

Strictly speaking, a partition of a graph in communities is a split of the graph
in clusters, with each vertex assigned to only one cluster.
The latter condition may be relaxed, as shown in Section~\ref{sec14}. 
Whatever the definition of community is, there is usually a large number of possible partitions.
It is then necessary to establish which partitions exhibit a
real community structure.
For that, one needs 
a {\it quality function}, i.e. a quantitative criterion to evaluate how good a partition is. The most popular quality function
is the modularity of Newman and Girvan~\cite{Newman:2004b}. It can be written in several ways, as
\begin{equation}
Q=\frac{1}{2m}\sum_{ij}\left(A_{ij}-\frac{k_ik_j}{2m}\right)\delta(C_i,C_j),
\label{eq:mod}
\end{equation}
where the sum runs over all pairs of vertices, $A$ is the adjacency matrix, $k_i$ the degree of vertex $i$
and $m$ the total number of edges of the graph. The element $A_{ij}$ of the adjacency matrix is $1$
if vertices $i$ and $j$ are connected, otherwise it is $0$.
The $\delta$-function yields one if vertices $i$ and $j$ are in the same
community, zero otherwise. Because of that, the only contributions to the sum come from vertex pairs belonging to the same
cluster: by grouping them together the sum over the vertex pairs can be rewritten as a sum over the modules
\begin{equation}
Q=\sum_{s=1}^{n_m}\Big[\frac{l_s}{m}-\left(\frac{d_s}{2m}\right)^2\Big].
\label{eq:mod1}
\end{equation}
Here, $n_m$ is the number of modules, $l_s$ the total number of edges joining vertices of module $s$ and 
$d_s$ the sum of the degrees of the vertices of $s$. In Eq.~\ref{eq:mod1}, the first term of each summand is the 
fraction of edges of the graph inside the module, whereas the second term represents the expected fraction of 
edges that would be there if the graph were a random graph with the same degree for each vertex.
In such a case,  
a vertex could be attached to any other vertex of the graph, and the probability of a connection between two vertices 
is proportional to the product of their degrees. So, for a vertex pair, the comparison between real
and expected edges is expressed by the corresponding summand of Eq.~\ref{eq:mod}.

Eq.~\ref{eq:mod1} embeds an implicit definition of community: a subgraph is a module if the number of edges inside it is larger 
than the expected number in modularity's null model. If this is the case, the vertices of the subgraph are more tightly 
connected than expected.
Basically, if each summand in Eq.~\ref{eq:mod1} is non-negative,
the corresponding subgraph is a module. Besides, the larger the difference between real and expected edges, the more ``modular''
the subgraph. So, large positive values of $Q$ are expected to indicate good partitions. The modularity of the whole graph, taken as a single
community, is zero, as the two terms of the only summand in this case are equal and opposite. 
Modularity is always smaller than one,
and can be negative as well. For instance, the partition in which each vertex is a community is always negative.
This is a nice feature of the measure, implying that, if there are no partitions with positive modularity,
the graph has no community structure. On the contrary, the existence of partitions with large negative 
modularity values may hint to the existence of 
subgroups with very few internal edges and many edges lying between them ({\it multipartite structure}). 

Modularity has been employed as quality function in many algorithms, like some of the divisive algorithms 
of Section~\ref{sec41}. In addition, modularity optimization is itself a popular method for community detection 
(see Section~\ref{sec42}). Modularity also allows to assess the stability of partitions~\cite{massen06} 
and to transform a graph into a smaller one by preserving its community structure~\cite{arenas07}.

However, there are some caveats on the use of the measure. The most important
concerns the value of modularity for a partition. For which values one can
say that there is a clear
community structure in a graph? The question is tricky: if two graphs have the same type of modular structure,
but different sizes, modularity will be larger for the larger graph. So, modularity values cannot be
compared for different graphs. Moreover, one would expect that partitions of random graphs
will have modularity values close to zero, as no community structure is expected there. Instead, 
it has been shown that partitions of random graphs may attain fairly large modularity values, as 
the probability that the distribution of edges on the vertices is locally 
inhomogeneous in specific realizations is not negligible~\cite{Guimera:2004}. 
Finally, a recent analysis has proved that modularity increases if subgraphs smaller than 
a characteristic size are merged~\cite{fortunato:2007}.
This fact represents a serious bias when one looks for communities via modularity optimization
and is discussed in more detail in Section~\ref{sec42}.

\subsection{Hierarchies}
\label{sec02}

\begin{figure}
\begin{center}
\includegraphics[width=6cm]{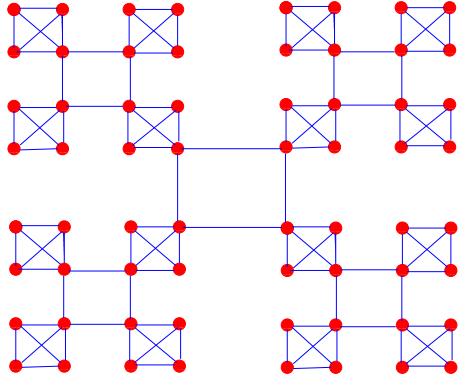}
\caption {\label{Figure2} Schematic example of a hierarchical graph. Sixteen modules with four vertices 
each are clearly organized in groups of four.}
\end{center}
\end{figure}

Graph vertices can have various levels of organization. Modules can display
an internal community structure, 
i.e. they can contain smaller modules, which can in turn include other modules, and so on. 
In this case one says that the graph is hierarchical (see Fig.~\ref{Figure2}).
For a clear classification of the vertices and
their roles inside a graph, it is important to find all modules of the graph as well as their hierarchy.

\begin{figure}
\begin{center}
\includegraphics[width=6cm]{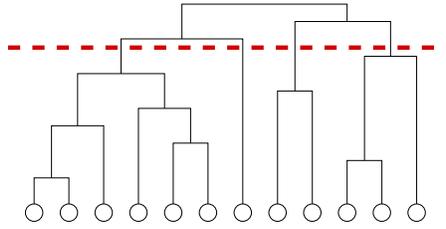}
\caption {\label{Figure3} A dendrogram, or hierarchical tree. Horizontal cuts correspond to partitions of the graph in communities.
Reprinted figure with permission 
from Newman MEJ, Girvan M, Physical Review E 69, 026113, 2004. Copyright 2004 by the Americal Physical Society.}
\end{center}
\end{figure}

A natural way to represent the hierarchical structure of a graph is to draw a {\it dendrogram}, like the one
illustrated in Fig.~\ref{Figure3}. Here, partitions of a graph with twelve vertices are shown. At the bottom,
each vertex is its own module. By moving upwards, groups of vertices are successively aggregated. Merges of 
communities are represented by horizontal lines. The uppermost level represents the whole graph as a single community. 
Cutting the diagram horizontally at some height, as shown in the figure (dashed line), 
displays one level of organization of the graph vertices. The diagram
is hierarchical by construction: each community belonging to a level is fully included in a community at a higher level.
Dendrograms are regularly used in sociology and biology. The technique of hierarchical clustering, described in Section~\ref{sec3},
lends itself naturally to this kind of representation.

\subsection{Overlapping Communities}
\label{sec14}

\begin{figure}
\begin{center}
\includegraphics[width=5cm]{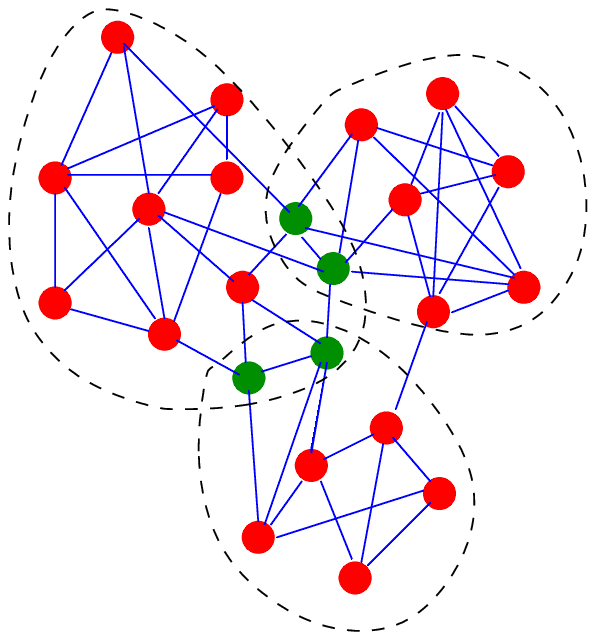}
\caption {\label{Figure4} Overlapping communities. In the partition highlighted by the dashed contours, 
the green vertices are shared between more groups.}
\end{center}
\end{figure}

As stated in Section~\ref{sec01}, in a partition each vertex is generally attributed only to one module.
However, vertices lying at the boundary between modules may be difficult to assign to one module or another,
based on their connections with the other vertices. In this case, it makes sense to consider such intermediate vertices as belonging
to more groups, which are then called {\it overlapping communities} (Fig.~\ref{Figure4}). 
Many real networks are characterized by a modular structure with sizeable overlaps between different clusters.
In social networks, people usually belong to more communities, according to their personal life and interests:
for instance a person may have tight relationships both with the people of its working environment and with other individuals   
involved in common free time activities. 

Accounting for overlaps is also a way to better exploit the  
information that one can derive from topology. Ideally, one could 
estimate the degree of participation of a vertex 
in different communities, which corresponds to the likelihood that the vertex belongs to the various groups.
Community detection algorithms, instead, often disagree in the
classification of periferal vertices of modules,
because they are forced to put them in a single cluster, which may be the wrong one.

The problem of community detection is so hard that very few algorithms consider
the possibility of having overlapping communities.  
An interesting method has been recently proposed by G. Palla et al.~\cite{Palla:2005} and is described in Section~\ref{sec45}.
For standard algorithms, the problem of identifying overlapping vertices could be addressed by 
checking for the stability of partitions against slight variations in the structure of the graph,
as described in~\cite{gfeller05}.

\section{Computer Science: Graph Partitioning}
\label{sec2}

\begin{figure}
\begin{center}
\includegraphics[width=5cm]{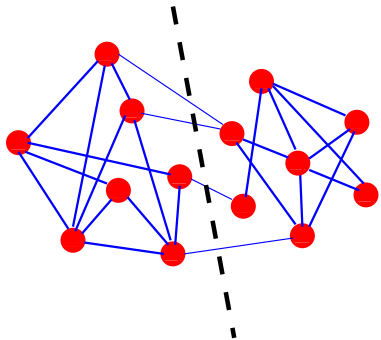}
\caption {\label{Figure5} Graph partitioning. The cut shows the partition in two groups of equal size.}
\end{center}
\end{figure}

The problem of graph partitioning consists in dividing the vertices in $g$ groups of predefined size, 
such that the number of edges lying between the groups is minimal.
The number of edges running between modules is called {\it cut size}.
Fig.~\ref{Figure5} presents the solution
of the problem for a graph with fourteen vertices, for $g=2$ and clusters of equal size. 

The specification of the number of modules of the partition is necessary.
If one simply imposed
a partition with the minimal cut size, and left the number of modules free, the solution would be trivial, corresponding to 
all vertices ending up in the same module, as this would yield a vanishing cut size. 

Graph partitioning is a fundamental issue in parallel computing, circuit partitioning and layout, and in the 
design of many serial algorithms, including techniques to solve partial differential equations and sparse
linear systems of equations.
Most variants of the graph partitioning problem are NP-hard, i.e. it is unlikely that the solution
can be computed in a time growing as a power of the graph size. There are however several algorithms that 
can do a good job, even if their solutions are not necessarily optimal~\cite{pothen}. Most algorithms 
perform a bisection of the graph, which is already a complex task. Partitions into more than two
modules are usually attained by iterative bisectioning. 

The {\it Kernighan-Lin algorithm}~\cite{kernighan} is one of the earliest 
methods proposed and is still frequently used, often in combination
with other techniques. The authors were motivated by the problem of partitioning electronic
circuits onto boards: the nodes contained in different boards need to be linked 
to each other with the least number of connections. The procedure is an
optimization
of a benefit function $Q$, which represents the difference between the number of edges inside 
the modules and the number of edges lying between them. The starting point is an initial 
partition of the graph in two clusters of the predefined size: such initial partition can be random 
or suggested by some information on the graph structure. Then, subsets consisting of 
equal numbers of vertices are swapped between the two groups, so that $Q$ has the maximal increase.
To reduce the risk to be trapped in local maxima of $Q$, the procedure includes some swaps that decrease the 
function $Q$. After a series of swaps with positive and negative gains, the partition with the largest value of 
$Q$ is selected and used as starting point of a new series of iterations. The Kernighan-Lin algorithm
is quite fast, scaling as $O(n^2)$ in worst-case time, $n$ being the number of vertices. 
The partitions found by the procedure are strongly
dependent on the initial configuration and other algorithms can do better. However, the method 
is used to improve on the partitions found through other techniques, by using them as starting
configurations for the algorithm.

Another popular technique is the {\it spectral bisection method}, which is based on the 
properties of the Laplacian matrix. The Laplacian matrix (or simply Laplacian) of a graph
is obtained from the adjacency matrix $A$ by placing on the diagonal the degrees of the vertices and by 
changing the signs of the other elements. The Laplacian has all non-negative eigenvalues and at least one 
zero eigenvalue, as the sum of the elements of each row and column of the matrix is zero.
If a graph is divided into $g$ connected components,
the Laplacian would have $g$ degenerate eigenvectors with eigenvalue zero and can be written in 
block-diagonal form, i.e. the vertices can be ordered in such a way that the Laplacian displays
$g$ square blocks along the diagonal, with entries different from zero, whereas all other elements vanish.
Each block is the Laplacian of the corresponding subgraph, so it has the trivial eigenvector with components
$(1,1,1,...,1,1)$. Therefore, there are $g$ degenerate eigenvectors with equal non-vanishing components in correspondence
of the vertices of a block, whereas all other components are zero. In this way, from the components of 
the eigenvectors one can identify the connected components of the graph.

If the graph is connected, but consists of $g$ subgraphs which are weakly linked to each other, the spectrum
will have one zero eigenvalue and $g-1$ eigenvalues which are close to zero. If the groups are two,
the second lowest eigenvalue will be close to zero and the corresponding eigenvector, also called
{\it Fiedler vector}, can be used to identify the two clusters as shown below.

Every partition of a graph with $n$ vertices in two groups can be represented by  
an index vector ${\bf s}$, whose component ${\bf s}_i$ is $+1$ if vertex $i$ is in one group and $-1$
if it is in the other group. The cut size $R$ of the partition of the graph in the two
groups can be written as
\begin{equation}
R=\frac{1}{4}{\bf s}^T{\bf Ls},
\label{eqr}
\end{equation}
where ${\bf L}$ is the Laplacian matrix and ${\bf s}^T$ the transpose of vector ${\bf s}$.
Vector ${\bf s}$ can be written as ${\bf s}=\sum_{i}a_i{\bf v}_i$, 
where ${\bf v}_i$, $i=1,...,n$ are the eigenvectors of the Laplacian. If ${\bf s}$ is properly normalized, then
\begin{equation}
R=\sum_i a_i^2\lambda_i,
\label{eqr1}
\end{equation}
where $\lambda_i$ is the Laplacian eigenvalue corresponding to eigenvector ${\bf v}_i$.
It is worth remarking that the sum contains at most $n-1$ terms, as the Laplacian has at least one zero eigenvalue.
Minimizing $R$ equals to the minimization of the sum on the right-hand side of Eq.~\ref{eqr1}.
This task is still very hard. However, if the second lowest eigenvector
$\lambda_2$ is close enough to zero, a good approximation of the minimum can be attained by choosing ${\bf s}$
parallel to the Fiedler vector ${\bf v}_2$: this would reduce the sum to $\lambda_2$, which is 
a small number. But the index vector cannot be perfectly parallel to ${\bf v}_2$ by construction, because all
its components are equal in modulus, whereas the components of ${\bf v}_2$ are not. The best one can do is to
match the signs of the components. So, one can set ${\bf s}_i=+1$ ($-1$) if ${\bf v}_2^i>0$ ($<0$).
It may happen that the sizes of the two corresponding groups do not match the predefined sizes one wishes to have.
In this case, if one aims at a split in $n_1$ and $n_2=n-n_1$ vertices, the best strategy is to order the components 
of the Fiedler vector from the lowest to the largest values and to put in one group the vertices corresponding 
to the first $n_1$ components from the top or the bottom, and 
the remaining vertices in the second group. If there is a discrepancy between $n_1$ and the number of positive or negative components 
of ${\bf v}_2$, this procedure yields two partitions: the better solution is the one that gives the smaller cut size.

The spectral bisection method is quite fast. The first eigenvectors of the Laplacian can be computed by using the Lanczos
method~\cite{lanczos}, that scales as $m/(\lambda_3-\lambda_2)$, where $m$ is the number of edges of the graph. If the eigenvalues
$\lambda_2$ and $\lambda_3$ are well separated, the running time of the algorithm is much shorter than the time required
to calculate the complete set of eigenvectors, which scales as $O(n^3)$. The method gives in general good partitions, that can be further
improved by applying the Kernighan-Lin algorithm. 

Other methods for graph partitioning include level-structure partitioning, the geometric algorithm, multilevel algorithms, etc.
A good description of these algorithms can be found in Ref.~\cite{pothen}.

Graph partitioning algorithms are not good for community detection, because it is necessary to provide as 
input both the number of groups and their size, about which in principle one knows nothing.
Instead, one would like an algorithm capable to produce this information
in its output. Besides, using iterative bisectioning 
to split the graph in more pieces is not a reliable procedure.

\section{Social Science: Hierarchical and K-Means Clustering}
\label{sec3}

In social network analysis, one partitions actors/vertices in clusters such that
actors in the same cluster are more similar between themselves than actors of different clusters. 
The two most used techniques to perform clustering analysis in sociology are
{\it hierarchical clustering} and {\it k-means clustering}.

The starting point of hierarchical clustering is the definition of a similarity measure between
vertices. After a measure is chosen, one computes the similarity for each pair of vertices, no matter 
if they are connected or not. At the end of this process, one is left with a 
new $n\times n$ matrix $X$, the similarity matrix. Initially, there are $n$ groups, each containing one of the vertices.
At each step, the two most similar groups are merged; the procedure continues until all vertices are in the same group. 

There are different ways to define the similarity between groups out of the matrix $X$.
In {\it single linkage clustering}, the similarity between two groups is the minimum element $x_{ij}$, with $i$ in one group 
and $j$ in the other. On the contrary, the maximum element $x_{ij}$ for vertices of different groups
is used in the procedure of {\it complete linkage clustering}. In {\it average linkage clustering} one has to  
compute the average of the $x_{ij}$.

The procedure can be better illustrated by means of dendrograms, like the one in Fig.~\ref{Figure3}. One should note that 
hierarchical clustering does not deliver a single partition, but a set of partitions.

There are many possible ways to define a similarity measure for the vertices based on the topology of the network.
A possibility is to define a distance between vertices, like
\begin{equation}
x_{ij}=\sqrt{\sum_{k\neq i,j}(A_{ik}-A_{jk})^2}.
\label{eqr2}
\end{equation}
This is a dissimilarity measure, based on the concept of structural equivalence. Two vertices are 
structurally equivalent if they have the same neighbours, even if they are not
adjacent themselves. If $i$ and $j$ are structurally equivalent,
$x_{ij}=0$. Vertices with large degree and different neighbours are considered very ``far'' from each other.
Another measure related to structural equivalence is the Pearson 
correlation between columns or rows of the adjacency matrix,
\begin{equation}
x_{ij}=\frac{\sum_k (A_{ik}-\mu_i)(A_{jk}-\mu_j)}{n\sigma_i\sigma_j},
\label{eqr3}
\end{equation}
where the averages $\mu_i=(\sum_jA_{ij})/n$ and the variances $\sigma_i=\sum_j (A_{ij}-\mu_i)^2$.

An alternative measure is the number of edge- (or vertex-) independent paths between two vertices. Independent
paths do not share any edge (vertex), and their number is related to the maximum flow that can be
conveyed between the two vertices under the constraint that each edge can carry only one unit of flow
(max-flow/min-cut theorem).
Similarly, one could consider all paths running between two vertices.
In this case, there is the problem that the total number of paths is infinite, but this can be avoided if 
one performs a weighted sum of the number of paths, where paths of length $l$ are weighted by 
the factor $\alpha^l$, with $\alpha<1$. So, the weights of long paths are exponentially suppressed and the sum converges.

Hierarchical clustering has the advantage that it does not require a preliminary knowledge on the number 
and size of the clusters. However, it does not provide a way to discriminate between the many 
partitions obtained by the procedure, and to choose that or those that better represent the community
structure of the graph. Moreover, the results of the method depend on the specific similarity measure adopted.
Finally, it does not correctly classify all vertices of a community, and in many 
cases some vertices are missed even if they have a central role in their clusters~\cite{newman04rev}. 

Another popular clustering technique in sociology is k-means clustering~\cite{macqueen}. Here, the number of clusters
is preassigned, say $k$. The vertices of the graph are embedded in a metric space, so that each vertex is a point and a distance measure is defined between
pairs of points in the space. 
The distance is a measure of dissimilarity between vertices.
The aim of the algorithm is to identify $k$ points in this space,
or {\it centroids}, so that each vertex is 
associated to one centroid and the sum of the distances of all vertices from their respective centroids
is minimal. To achieve this, one starts from an initial distribution of centroids such that they are 
as far as possible from each other. In the first iteration, each vertex is assigned to the nearest centroid.
Next, the centers of mass of the $k$ clusters are estimated and become a new set of 
centroids, which allows for a new classification of the vertices, and so on. 
After a sufficient number of iterations, the positions of the centroids are stable,
and the clusters do not change any more.
The solution found is not necessarily optimal,
as it strongly depends on the initial choice of the centroids.
The result can be improved by performing more runs starting from different initial conditions.

The limitation of k-means clustering is the same
as that of the graph partitioning algorithms: the number of clusters
must be specified at the beginning, the method is not able to derive it.
In addition, the embedding in a metric space can be natural for some graphs,
but rather artificial for others.

\section{New Methods}
\label{sec4}

From the previous two sections it is clear that traditional approaches to derive graph partitions
have serious limits. The most important problem is the need to provide the algorithms with  
information that one would like to derive from the algorithms themselves, like the number of clusters and their size.
Even when these inputs are not necessary, like in hierarchical clustering, there is the question of estimating
the goodness of the partitions, so that one can pick the best one. For these reasons, 
there has been a major effort in the last years to devise algorithms capable of extracting a complete 
information about the community structure of graphs. These methods can be grouped in different categories.

\subsection{Divisive Algorithms}
\label{sec41}

A simple way to identify communities in a graph is to detect the edges that connect vertices 
of different communities and remove them, so that the clusters get disconnected from each other. This is the
philosophy of divisive algorithms. The crucial point is to find a property of intercommunity edges that could allow for their identification. 
Any divisive method delivers  many partitions, which are by construction
hierarchical, so that they can be represented with dendrograms. 

\noindent {\it Algorithm of Girvan and Newman}. The most popular algorithm is that proposed by Girvan 
and Newman~\cite{Girvan:2002}. The method is also historically important, because it marked the beginning
of a new era in the field of community detection. Here edges are selected according to the values of 
measures of {\it edge centrality}, estimating the importance of edges
according to some property or process running on the graph. 
The steps of the algorithm are:
\begin{enumerate}
\item{Computation of the centrality for all edges;}
\item{Removal of edge with largest centrality;}
\item{Recalculation of centralities on the running graph;}
\item{Iteration of the cycle from step 2.}
\end{enumerate}
Girvan and Newman 
focused on the concept of {\it betweenness}, which is a variable expressing the frequency of the participation 
of edges to a process. They considered three alternative definitions: 
edge betweenness, current-flow betweenness and random walk betweenness.

\begin{figure}
\begin{center}
\includegraphics[width=6cm]{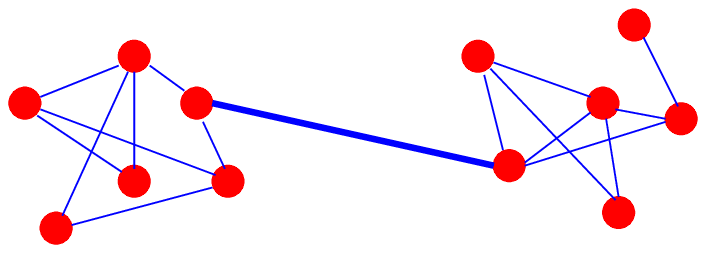}
\caption {\label{Figure6} Edge betweenness is highest for edges connecting communities. In the figure, the thick
edge in the middle has a much higher betweenness than all other edges, because all shortest paths connecting vertices of the
two communities run through it.}
\end{center}
\end{figure}

Edge betweenness is the number of shortest paths between all vertex pairs that run along the edge.
It is an extension to edges of the concept of site betweenness, introduced by Freeman in 1977~\cite{freeman}.
It is intuitive that intercommunity edges have a large value of the edge betweenness, because
many shortest paths connecting vertices of different communities will pass through them (Fig.~\ref{Figure6}).
The betweenness of all edges of the graph can be calculated in a time that scales as $O(mn)$, with 
techniques based on breadth-first-search~\cite{Newman:2004b,brandesSB}. 

Current-flow betweenness is defined by considering the graph a resistor network, with edges having 
unit resistance. If a voltage difference is applied 
between any two vertices, each edge carries some amount of current, that can be calculated by solving 
Kirchoff's equations. The procedure is repeated for all possible vertex pairs: the current-flow
betweenness of an edge is the average value of the current 
carried by the edge. Solving Kirchoff's equations
requires the inversion of an $n\times n$ matrix, which can be done in a time $O(n^3)$ for a sparse matrix.

The random-walk betweenness of an edge says how frequently a random walker running on the graph goes across the edge. 
We remind that a random walker moving from a vertex follows each edge with equal probability.
A pair of vertices is chosen at random, $s$ and $t$. The walker starts at $s$ and keeps moving until
it hits $t$, where it stops. One computes the probability that each edge was crossed by the walker, and averages over all
possible choices for the vertices $s$ and $t$. The complete calculation requires a time $O(n^3)$ on a sparse graph.
It is possible to show that this measure is equivalent to 
current-flow betweenness~\cite{newman05RW}.

Calculating edge betweenness is much faster than current-flow or random walk betweenness 
($O(n^2)$ versus $O(n^3)$ on sparse graphs).
In addition, in practical applications the Girvan-Newman algorithm with edge
betweenness gives better results than adopting the other centrality measures.
Numerical studies show that the
recalculation step 3 of Girvan-Newman algorithm is essential to detect meaningful communities.
This introduces an additional factor $m$ in the running time of the algorithm: consequently, the edge betweenness
version scales as $O(m^2n)$, or
$O(n^3)$ on a sparse graph. Because of that, the algorithm is quite slow, and applicable to
graphs with up to $n\sim 10000$ vertices, with current computational resources. In the original version of 
Girvan-Newman's algorithm~\cite{Girvan:2002}, the authors had to deal with the whole hierarchy of partitions, as they 
had no procedure to say which partition is the best. In a successive refinement~\cite{Newman:2004b}, they 
selected the partition with the largest value of modularity (see Section~\ref{sec01}), a criterion that has been
frequently used ever since. There have been countless applications of the Girvan-Newman method: the algorithm  
is now integrated in well known libraries of network analysis programs.

\noindent {\it Algorithm of Tyler et al.}. Tyler, Wilkinson and Huberman proposed a modification of the Girvan-Newman
algorithm, to improve the speed of the calculation~\cite{tyler}. The modification consists in
calculating the contribution to edge betweenness only from a limited number of vertex pairs, chosen at random,
deriving a sort of Monte Carlo estimate. The procedure induces statistical errors in the values of 
the edge betweenness. As a consequence, the partitions are in general different for different
choices of the sampling pairs of vertices. However, the authors showed that, by repeating the
calculation many times, the method gives
good results, with a substantial gain of computer time. In practical examples, only 
vertices lying at the boundary between communities may not be clearly classified, and be assigned 
sometimes to a group, sometimes to another. 
The method has been applied to a network of people corresponding through email~\cite{tyler} and to networks of 
gene co-occurrences~\cite{wilkinson}.

\noindent {\it Algorithm of Fortunato et al.}. An alternative measure of centrality for edges is 
information centrality. It is based on the concept of efficiency~\cite{latora}, which estimates how easily information
travels on a graph according to the length of shortest paths between
vertices. The information centrality of an edge is the variation of the efficiency of the graph
if the edge is removed. In the algorithm by Fortunato, Latora and Marchiori~\cite{FLM}, edges are removed
according to decreasing values of information centrality. The method is analogous to that of Girvan and Newman,
but slower, as it scales as $O(n^4)$ on a sparse graph. On the other hand, it gives a better classification of vertices when 
communities are fuzzy, i.e. with a high degree of interconnectedness.

\noindent {\it Algorithm of Radicchi et al.}. Because of the high density of edges within communities,
it is easy to find loops in them, i.e. closed non-intersecting paths. On the contrary, edges lying between communities will hardly be 
part of loops. Based on this intuitive idea, Radicchi et al. proposed a new measure, the edge clustering coefficient, such that 
low values of the measure are likely to correspond to intercommunity edges~\cite{radicchi}. The edge clustering coefficient generalizes to edges the
notion of clustering coefficient introduced by Watts and Strogatz for vertices~\cite{watts98}.
The latter is the number of triangles including a vertex divided by the number of possible triangles that can be formed.
The edge clustering coefficient is the number of loops of length $g$
including the edge divided by the number of possible cycles. Usually, loops of length $g=3$ or $4$ 
are considered. At each iteration, the edge with smallest clustering coefficient
is removed, the measure is recalculated again, and so on. The procedure stops when all clusters obtained are LS-sets or 
``weak'' communities (see Section~\ref{seccomm}).
Since the edge clustering coefficient is a local measure, involving
at most an extended neighbourhood of the edge, it can be calculated very quickly. The running time of the algorithm to completion
is $O(m^4/n^2)$, or $O(n^2)$ on a sparse graph, so it is much shorter than the running time of the Girvan-Newman method.
On the other hand, the method may give poor results when the graph has few loops, as it happens in several non-social networks.
In this case, in fact, the edge clustering coefficient is small and fairly similar for all edges, and the algorithm 
may fail to identify the bridges between communities.

\subsection{Modularity Optimization}
\label{sec42}

If Newman-Girvan modularity $Q$ (Section~\ref{sec01}) is a good indicator of the quality of partitions, 
the partition corresponding to its maximum value 
on a given graph should be the best, or at least a very good one. 
This is the main motivation for modularity maximization,
perhaps the most popular class of methods to detect communities in graphs.
An exhaustive optimization of $Q$ is 
impossible, due to the huge number of ways in which it is possible to partition a graph, even when the latter is small. 
Besides, the true maximum is out of reach, as it has been recently proved that modularity optimization is 
an NP-hard problem~\cite{brandes}, so it is probably impossible to find the solution in a time 
growing polynomially with the size of the graph.
However, there are currently several algorithms able to find fairly good approximations of 
the modularity maximum in a reasonable time.

\noindent {\it Greedy techniques}. The first algorithm devised to maximize modularity 
was a greedy method of Newman~\cite{Newman:2004c}.
It is an agglomerative method, where groups of vertices are successively joined to form larger communities such that modularity 
increases after the merging. One starts from $n$ clusters, each containing a single vertex. Edges are not initially 
present, they are added one by one during the procedure. However, modularity is always 
calculated from the full topology of the graph,
since one wants to find its partitions. Adding a first edge to the set of disconnected 
vertices reduces the number of groups from $n$ to
$n-1$, so it delivers a new partition of the graph. The edge is chosen such that this partition gives the maximum 
increase of modularity with respect to the previous configuration. 
All other edges are added based on the same principle. If the insertion of an edge does not change the
partition, i.e. the clusters are the same, modularity stays the same.
The number of partitions
found during the procedure is $n$, each with a different number of clusters, from $n$ to $1$. 
The largest value of modularity in this subset of partitions is the approximation of the modularity 
maximum given by the algorithm. The update of the modularity value at each iteration step can be performed
in a time $O(n+m)$, so the algorithm runs to completion in a time $O((m+n)n)$, or $O(n^2)$
on a sparse graph, which is fast. In a later paper by Clauset et al.~\cite{Clauset:2004}, it was shown that the 
calculation of modularity during the procedure can be performed 
much more quickly by use of max-heaps, special data structures created using a binary tree.
By doing that, the algorithm scales as $O(md\log n)$, where $d$ is the depth of the dendrogram describing 
the successive partitions found during the execution of the algorithm, which 
grows as $\log n$ for graphs with a strong hierarchical structure. For those graphs, the running time of the
method is then $O(n\log^2n)$, which allows to analyse the community structure
of very large graphs, up to $10^7$ vertices. The greedy algorithm is currently the only algorithm that can be used
to estimate the modularity maximum on such large graphs.
On the other hand, the approximation 
it finds is not that good, as compared with other techniques. The accuracy of the
algorithm can be considerably improved if one accounts for the size of the groups to be merged~\cite{danon06},
or if the hierarchical agglomeration is started from some good intermediate configuration, rather than from
the individual vertices~\cite{pujol06}.

\noindent {\it Simulated annealing}. Simulated annealing~\cite{simann} is a
probabilistic procedure for global optimization used in different fields and problems.
It consists in performing an exploration of the space of possible states, looking for the
global optimum of a function $F$, say its maximum. Transitions from one state to another occur with probability
$1$ if $F$ increases after the change, otherwise
with a probability $\exp(\beta\Delta F)$, where $\Delta F$ is the decrease of the function and 
$\beta$ is an index of stochastic noise, a sort of inverse temperature, which increases after
each iteration. 
The noise reduces the risk that the system gets trapped in local optima. At some stage, 
the system converges to a stable state, which can be an arbitrarily good approximation of the
maximum of $F$, depending on how many states were explored and how slowly $\beta$ is varied.
Simulated annealing was first employed for modularity optimization by R.~Guimer\'a et al.~\cite{Guimera:2004}.
Its standard implementation combines two types of ``moves'': local moves, where a single 
vertex is shifted from one cluster to another, taken at random; global moves, consisting 
of merges and splits of communities. In practical applications, one typically combines $n^2$ local moves with 
$n$ global ones in one iteration. The method can potentially come very close to the 
true modularity maximum, but it is slow.
Therefore, it can be used for small graphs, with up to about $10^4$ vertices.
Applications include studies of potential energy landscapes~\cite{massen05} and of metabolic 
networks~\cite{Guimera:2005}.

\noindent{\it Extremal optimization}. Extremal optimization is a heuristic search procedure proposed
by Boettcher and Percus~\cite{boettcher}, in order to achieve an accuracy comparable with simulated
annealing, but with a substantial gain in computer time. It is based on the optimization of 
local variables, expressing the contribution of each unit of the system to the global function at study.
This technique was used for modularity optimization by Duch and Arenas~\cite{Duch:2005}.
Modularity can be indeed written as a sum over the vertices: the local modularity of a vertex
is the value of the corresponding term in this sum. A fitness measure for each vertex is obtained
by dividing the local modularity of the vertex by its degree. One starts from a random partition of the graph 
in two groups. At each iteration, the vertex with the lowest fitness is shifted to the other cluster.
The move changes the partition, so the local fitnesses need to be recalculated. The process continues 
until the global modularity $Q$ cannot be improved any more by the procedure. At this stage, each cluster is 
considered as a graph on its own and the procedure is repeated, as long as $Q$ increases for the 
partitions found. The algorithm finds an excellent approximation of the modularity maximum
in a time $O(n^2\log n)$, so it represents a good tradeoff between accuracy and speed.

\noindent {\it Spectral optimization}. Modularity can be
optimized using the eigenvalues and eigenvectors of a special matrix, the modularity matrix $B$, whose elements are
\begin{equation}
B_{ij}=A_{ij}-\frac{k_ik_j}{2m},
\label{eqr4}
\end{equation}
where the notation is the same used in Eq.~\ref{eq:mod}. The method~\cite{Newman:2006,Newman:2006b} is analogous to
spectral bisection, described in Section~\ref{sec2}. The difference is that here 
the Laplacian matrix is replaced by the modularity matrix. Between $Q$ and $B$ there is the 
same relation as between $R$ and $T$ in Eq.~\ref{eqr}, so modularity can be written as a weighted 
sum of the eigenvalues of $B$, just like Eq.~\ref{eqr1}. 
Here one has to look for the eigenvector of $B$ with largest eigenvalue, ${\bf u}_1$,
and group the vertices according to the signs of the components of ${\bf u}_1$, just like in Section~\ref{sec2}. The Kernighan-Lin algorithm can then be used to improve the result.
The procedure is repeated for each of the clusters separately, and the number of communities
increases as long as modularity does.   
The advantage over spectral bisection is that it is not necessary to specify
the size of the two groups, because it is determined by taking the partition with largest modularity.
The drawback is similar as for spectral bisection, i.e. the algorithm gives the best results 
for bisections, whereas it is less accurate when the number of communities is larger than two.
The situation could be improved by using the other eigenvectors with positive eigenvalues of the modularity matrix.
In addition, the eigenvectors with the most negative eigenvalues are important to detect 
a possible multipartite structure of the graph, as they give the most relevant 
contribution to the modularity minimum.
The algorithm typically runs in a time $O(n^2\log n)$ for a sparse graph, when one computes only the first 
eigenvector, so it is faster than extremal optimization, and slightly more accurate, especially
for large graphs.

Finally, some general remarks on modularity optimization and its reliability.
A large value for the modularity maximum does not necessarily mean that a graph has community structure.
Random graphs can also have partitions with large modularity values, 
even though clusters are not explicitly built in~\cite{Guimera:2004, Reichardt07}. 
Therefore, the modularity maximum
of a graph reveals its community structure only if it is appreciably larger than the modularity maximum
of random graphs of the same size~\cite{Bornholdt:2006}.
\begin{figure}
\begin{center}
\includegraphics[width=6cm]{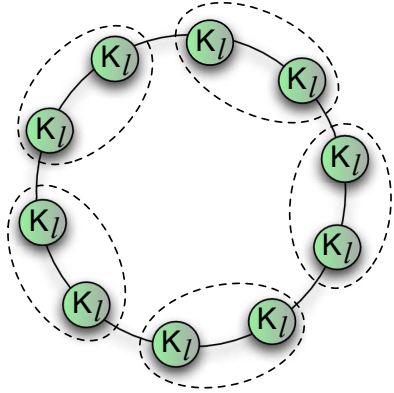}
\caption {\label{Figure7} Resolution limit of modularity optimization. The natural community
structure of the graph, represented by the individual cliques (circles), is not recognized by
optimizing modularity, if the cliques are smaller than a scale depending on the size of the graph. 
Reprinted figure with permission from Fortunato S, Barth\'elemy M, Proceedings of the National Academy of Science of the USA, 104, 36
(2007). Copyright 2007 from the National Academy of Science of the USA.}
\end{center}
\end{figure}

In addition, one assumes that the modularity maximum delivers the ``best'' partition of the network in communities.
However, this is not always true~\cite{fortunato:2007}. In the 
definition of modularity (Eq.~\ref{eq:mod1}) the graph is compared with a random version of it, that keeps the
degrees of its vertices. If groups of vertices in the graphs are more tightly connected than they would be
in the randomized graph, modularity optimization would consider them as parts of the same module. 
But if the groups have less than $\sqrt{m}$ internal edges, 
the expected number of edges running between them in modularity's null model
is less than one, and a single interconnecting edge would cause the merging of the two groups in the optimal partition.
This holds for every density of edges inside the groups, even in the limit case in which all vertices 
of each group are connected to each other, i.e. if the groups are cliques.
In Fig.~\ref{Figure7} a graph is made out of $n_c$ identical cliques,
with $l$ vertices each, connected by single edges.
It is intuitive to think that the modules of the best partition are the single cliques: instead,
if $n_c$ is larger than about $l^2$, modularity would be higher for the partition in which 
pairs of consecutive cliques are parts of the same module (indicated by the dashed lines in the figure).
The problem holds for a wide class of possible null models~\cite{kumpula:2007}.
Attempts 
have been made to solve it within the modularity framework~\cite{arenas07b,ruan07,kumpula:2007b}.

Modifications of the measure have also been suggested. Massen and Doye proposed a 
slight variation of modularity's null model~\cite{massen05}: 
it is still a graph with the same degree sequence as the original,
and with edges rewired at random among the vertices, but one imposes the additional constraint that
there can be neither multiple edges between a pair of vertices nor edges joining a vertex with itself (self-edges).
Muff, Rao and Caflisch remarked that modularity's null model implicitly assumes that each vertex could be attached
to any other, whether in real cases a cluster is usually connected to few other clusters~\cite{Muff:2005}. 
Therefore, they proposed a local version of modularity, in which the expected number of edges within 
a module is not calculated with respect to the full graph, but considering just a portion of it, namely the subgraph 
including the module and its neighbouring modules.

\subsection{Spectral Algorithms}
\label{sec43}

As discussed above, spectral properties of graph matrices are frequently used to find 
partitions. Traditional methods are in general unable to predict the number and size of the clusters,
which instead must be fed into the procedure. Recent algorithms, reviewed below, are more powerful.

\begin{figure}
\begin{center}
\includegraphics[width=6cm]{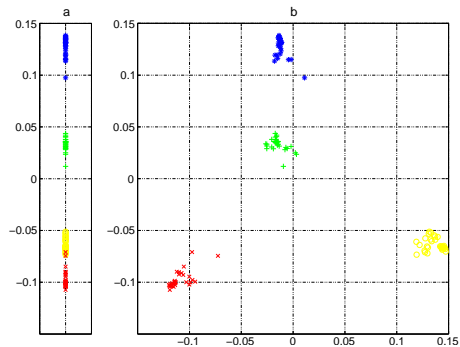}
\caption {\label{Figure8} Spectral algorithm by Donetti and Mu\~noz. Vertex $i$ is represented by the values of
the $i$th components of Laplacian eigenvectors. In this example, the graph has an ad-hoc division in four communities,
indicated by the colours. The communities are better separated in two dimensions (b) than in one (a). 
Reprinted figure with permission from Donetti L, Mu\~noz MA, Journal of Statistical Mechanics: Theory and Experiment, P10012 (2004).
Copyright 2004 by the Institute of Physics.}
\end{center}
\end{figure}

\noindent {\it Algorithm of Donetti and Mu\~noz}. An elegant method based on the eigenvectors of 
the Laplacian matrix has been devised by Donetti and Mu\~noz~\cite{donetti04}. The idea is simple: 
the values of the eigenvector components are close for vertices in the same community, so one can use them 
as coordinates to represent vertices as points in a metric space. So, if one uses $M$ eigenvectors,
one can embed the vertices in an $M$-dimensional space. Communities appear as groups of points well separated
from each other, as illustrated in Fig.~\ref{Figure8}. The separation is the more visible, the larger the number
of dimensions/eigenvectors $M$.
The space points are grouped in communities by hierarchical clustering (see Section~\ref{sec3}).
The final partition is the one with largest modularity.
For the similarity measure between vertices, Donetti and Mu\~noz used both the Euclidean distance
and the angle distance. The angle distance between two points 
is the angle between the vectors going from the origin of the
$M$-dimensional space to either point. Applications show that the best results are obtained with complete-linkage
clustering. The algorithm runs to completion in a time $O(n^3)$, which is not fast. Moreover,
the number $M$ of eigenvectors that are needed to have a clean separation of the clusters is 
not known {\it a priori}.

\noindent{\it Algorithm of Capocci et al.}. Similarly to Donetti and Mu\~noz, Capocci et al.
used eigenvector components to identify communities~\cite{capocci04}. In this case the eigenvectors are those of the
{\it normal matrix}, that is derived from the adjacency matrix by dividing each row by the sum of its elements.
The eigenvectors can be quickly calculated by performing a constrained optimization
of a suitable cost function. A similarity matrix is built by calculating the 
correlation between eigenvector components: the similarity between vertices $i$ and $j$ is the
Pearson correlation coefficient between their corresponding eigenvector components, where the averages
are taken over the set of eigenvectors used. The method can be extended to directed 
graphs. It is useful to estimate vertex similarities, however it does
not provide a well-defined partition of the graph.

\noindent{\it Algorithm of Wu and Huberman}. A fast algorithm by Wu and Huberman
identifies communities based on the properties of resistor networks~\cite{wu04}. 
It is essentially a method for bisectioning graph, similar to spectral bisection,   
although partitions in an arbitrary number of communities can be obtained by iterative applications.
The graph is transformed into a resistor network where each edge has unit resistance.
A unit potential difference is set between two randomly chosen vertices. The idea is that, if 
there is a clear division in two communities of the graph, there will be a visible
gap between voltage values for vertices at the borders between the clusters.
The voltages are calculated by solving Kirchoff's equations: an exact resolution would be too 
time consuming, but it is possible to find a reasonably good approximation in a linear time for a sparse graph with
a clear community structure, so the more time consuming part of the algorithm is the sorting of the voltage values, which
takes time $O(n\log n)$. Any possible vertex pair can be chosen to set the initial potential difference, so the 
procedure should be repeated for all possible vertex pairs. The authors showed that this is not necessary, and that
a limited number of sampling pairs is sufficient to get good results, so the algorithm scales as $O(n\log n)$ and is very fast.
An interesting feature of the method is that it can quickly find the natural community of any vertex,
without determining the complete partition of the graph. For that, one uses the vertex
as source voltage and places the sink at an arbitrary vertex.
The same feature is present 
in an older algorithm by Flake et al.~\cite{Flake:2002}, where one uses max-flow instead of current flow.

Previous works have shown that also the eigenvectors of the {\it transfer matrix} ${\bf T}$ can be used 
to extract useful information on community structure~\cite{eriksen03,simonsen04}. The element $T_{ij}$ of the 
transfer matrix is $1/k_j$ if $i$ and $j$ are neighbours, where $k_j$ is the degree
of $j$, otherwise it is zero. The transfer matrix rules the process of diffusion on graphs.

\subsection{Dynamic Algorithms}
\label{sec44}

This Section describes methods employing processes running on the graph,
focusing on spin-spin interactions, random walk and synchronization.

\noindent {\it Q-state Potts model}. The Potts model is among the most popular models in 
statistical mechanics~\cite{wu82}. It describes a system of spins that can be in $q$ 
different states. The interaction is ferromagnetic, i.e. 
it favours spin alignment, so at zero temperature all spins are in the same 
state. If antiferromagnetic interactions are also present, the ground state of the system may 
not be the one where all spins are aligned, but a state where different spin values coexist, in homogeneous clusters.
If Potts spin variables are assigned to the vertices of a graph with community structure, and the interactions 
are between neighbouring spins, it is likely that the topological clusters could be recovered from 
like-valued spin clusters of the system, as there are many more interactions inside communities than outside.
Based on this idea, inspired by an earlier paper by Blatt, Wiseman and Domany~\cite{blatt96}, Reichardt and Bornholdt
proposed a method to detect communities that maps the graph onto a $q$-Potts model with nearest-neighbours 
interactions~\cite{Reichardt04}. The Hamiltonian of the model, i.e. its energy, is the sum of two 
competing terms, one favoring spin alignment, one antialignment. The relative weight of these two terms
is expressed by a parameter $\gamma$, which is usually set to the value of the 
density of edges of the graph. The goal is to find the ground state of the system,
i.e. to minimize the energy. This can be done with simulated annealing~\cite{simann}, starting from 
a configuration where spins are randomly assigned to the vertices and the number of states $q$ is very high.
The procedure is quite fast and the results do not depend on $q$. The method also allows
to identify vertices shared between communities, from the comparison of partitions corresponding to 
global and local energy minima. More recently, Reichardt and Bornholdt derived a general framework~\cite{Reichardt06},
in which detecting community structure is equivalent to finding the ground state of a $q$-Potts model spin glass~\cite{parisi}.  
Their previous method and modularity optimization are recovered as special cases. Overlapping communities can be
discovered by comparing partitions with the same (minimal) energy, and 
hierarchical structure can be investigated by tuning a parameter acting on the density of edges 
of a reference graph without community structure.

\noindent {\it Random walk}. Using random walks to find communities comes from the idea that
a random walker spends a long time inside a community due to the high density of edges and 
consequent number of paths that could be followed. Zhou used random walks to define a distance 
between pairs of vertices~\cite{zhou03}: the distance between $i$ and $j$ is the average number of edges that a
random walker has to cross to reach $j$ starting from $i$. Close vertices are likely to belong to
the same community. The global attractor of a vertex $i$ is the closest
vertex to $i$, whereas the local attractor of $i$ is its closest neighbour. 
Two types of communities are defined, 
according to local or global attractors: a vertex $i$ has to be put in the same 
community of its attractor and of all other vertices for which $i$ is an attractor. Communities 
must be minimal subgraphs, i.e. they cannot include smaller subgraphs which are communities according
to the chosen criterion. Applications to real and artificial networks show that the method can find meaningful partitions.  
In a successive paper~\cite{zhou03b}, Zhou introduced a measure of dissimilarity between vertices based on 
the distance defined above. The measure resembles the definition of distance based on structural equivalence of 
Eq.~\ref{eqr2}, where the elements of the adjacency matrix are replaced by the corresponding distances.
Graph partitions are obtained with a divisive procedure that, starting from the graph as a single community,
performs successive splits based on the criterion that vertices in the same cluster must be less dissimilar than
a running threshold, which is decreased during the process. The hierarchy of partitions derived by the method
is representative of actual community structures for several real and artificial graphs.
In another work~\cite{zhou04}, Zhou and Lipowsky defined distances with biased random walkers, where the bias is
due to the fact that walkers move preferentially towards vertices sharing a large number of neighbours
with the starting vertex. A different distance measure between vertices based on
random walks was introduced by Latapy and Pons~\cite{latapy05}. The distance is calculated from the
probabilities that the random walker moves from a vertex to another in a fixed number of steps. Vertices are 
then grouped into communities through hierarchical clustering. The method is quite fast, 
running to completion in a time $O(n^2\log n)$ on a sparse graph.

\begin{figure}
\begin{center}
\includegraphics[width=7cm,angle=270]{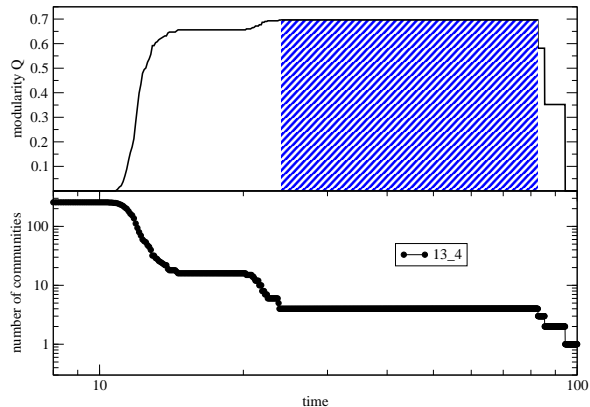}
\caption {\label{Figure9} Number of clusters of synchronized Kuramoto oscillators as a function of time for 
a hierarchical graph. The two levels of community structure are revealed by the plateaus in the figure, which indicate
the stability of those configurations. The top diagram shows the values of Newman-Girvan modularity $Q$ 
for the corresponding partitions. The shadowed area highlights the partition with largest modularity. Reprinted
figure with permission from Arenas A, D\'iaz-Guilera A, European Physical Journal ST 143, 19 (2007). Copyright 2007 by
EDP Sciences.}
\end{center}
\end{figure}

\noindent{\it Synchronization}. Synchronization is another promising dynamic process to reveal communities
in graphs. 
If oscillators are placed at the vertices, with initial random phases, and have nearest-neighbour interactions, 
oscillators in the same community synchronize first, whereas a full synchronization requires a longer time.
So, if one follows the time evolution of the process, states with synchronized clusters of vertices 
can be quite stable and long-lived, so they can be easily recognized.
This was first shown by Arenas, D\'iaz-Guilera and P\'erez-Vicente~\cite{arenas06}. They used 
Kuramoto oscillators~\cite{kuramoto}, which are coupled two-dimensional vectors endowed with a proper 
frequency of oscillations. If the interaction coupling exceeds a threshold, the dynamics leads to
synchronization. Arenas et al. showed that the time evolution of the system reveals some intermediate 
time scales, corresponding to topological scales of the graph, i.e. to different levels of organization of the 
vertices. 
%Hierarchical community structure can be revealed in this way. 
Hierarchical community structure can be revealed in this way (Fig.~\ref{Figure9}). 
Based on the same principle, Boccaletti et al. designed a community detection method based on synchronization~\cite{boccaletti07}.
The synchronization dynamics is a variation of Kuramoto's model, the opinion changing rate (OCR) model~\cite{pluchino05}.
The evolution equations of the model are solved for decreasing values of a parameter that tunes the 
strength of the interaction coupling between neighbouring vertices. In this way, different partitions
are recovered: the partition with the largest value of modularity is chosen.
The algorithm scales in a time $O(mn)$, or $O(n^2)$ on sparse graphs, and gives good results on practical examples.
However, synchronization-based algorithms may not be reliable when communities are very different in size.

\subsection{Clique Percolation}
\label{sec45}

\begin{figure}
\begin{center}
\includegraphics[width=7cm]{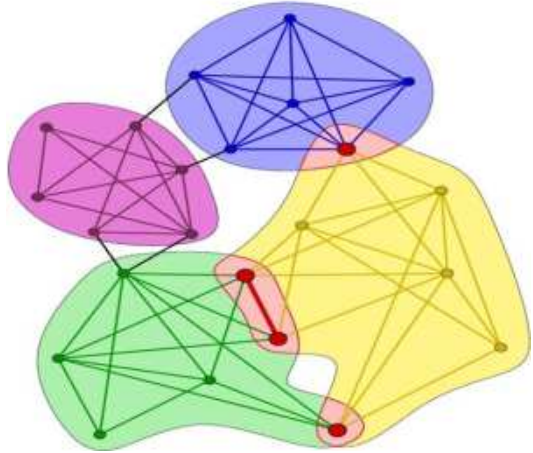}
\caption {\label{Figure10} Clique Percolation Method. The example shows communities spanned by 
adjacent 3-cliques (triangles). Overlapping vertices are shown by the bigger dots. Reprinted figure with 
permission from Palla G, Der\'enyi I, Farkas I and Vicsek T, Nature 435, 814 (2005). Copyright 2005 by the Nature
Publishing Group.}
\end{center}
\end{figure}

In most of the approaches examined so far, communities have been characterized and discovered, 
directly or indirectly, by some global property of the graph, like betweenness, modularity, etc.,  
or by some process that involves the graph as a whole, like random walks, synchronization, etc.
But communities can be also interpreted as a form of local organization of the graph, so they could be 
defined from some property of the groups of vertices themselves, regardless of the rest of the graph.
Moreover, very few of the algorithms presented so far are able to deal with the problem of overlapping communities
(Section~\ref{sec14}). A method that accounts both for the locality of the community definition
and for the possibility of having overlapping communities is the Clique Percolation Method (CPM)
by Palla et al.~\cite{Palla:2005}. It is based on the concept that the internal edges of community are likely
to form cliques due to their high density. On the other hand, it is unlikely that intercommunity edges
form cliques: this idea was already used in the divisive method of Radicchi et al. (see Section~\ref{sec41}).
Palla et al. define a $k$-clique as a complete graph with $k$ vertices.
Notice that this definition is different from the definition of
$n$-clique (see Section~\ref{seccomm}) used in social science.
If it were possible for a clique to move on a graph, in some way, it would probably get trapped
inside its original community, as it could not cross the bottleneck formed by the intercommunity edges.
Palla et al. introduced a number of concepts to implement this idea.
Two $k$-cliques are {\it adjacent} if
they share $k-1$ vertices. The union of adjacent $k$-cliques is called {\it $k$-clique chain}.
Two $k$-cliques are connected if they are part of a $k$-clique chain. Finally, a {\it $k$-clique
community} is the largest connected subgraph obtained by the union of a $k$-clique and 
of all $k$-cliques which are connected to it. Examples of $k$-clique communities are shown in Fig.~\ref{Figure10}.
One could say that a $k$-clique community is identified by making  
a $k$-clique ``roll'' over adjacent $k$-cliques, where rolling 
means rotating a $k$-clique about the $k-1$ vertices it shares with any adjacent $k$-clique.
By construction, $k$-clique communities can share vertices, so they can be overlapping.
There may be vertices belonging to non-adjacent $k$-cliques, which could be reached 
by different paths and end up in different clusters. In order to find $k$-clique communities,
one searches first for maximal cliques, a task that is known to require a running time
that grows exponentially with the size of the graph. However, the authors found that, for 
the real networks they analyzed, the procedure is quite fast, allowing to analyze graphs
with up to $10^5$ vertices in a reasonably short time. The actual scalability of the algorithm 
depends on many factors, and cannot be expressed in closed form. The algorithm has been extended 
to the analysis of weighted~\cite{palla07} and directed~\cite{palla07b} graphs. 
It was recently used to study the evolution of community structure in social networks~\cite{palla07c}.
A special software, called
{\it CFinder}, based on the CPM, has been designed by Palla and coworkers and is 
freely available. 
The CPM has the same limit as the algorithm of Radicchi et al.: 
it assumes that the graph has a large number of cliques, so it may fail to give meaningful partitions
for graphs with just a few cliques, like technological networks.
 
\subsection{Other Techniques}

This Section describes some algorithms that do not fit in the previous categories, although 
some overlap is possible.

\noindent {\it Markov Cluster Algorithm (MCL)}. This method, invented by van Dongen~\cite{vandongen},
simulates a peculiar process of flow diffusion in a graph. 
One starts from the {\it stochastic matrix} of the graph, which is obtained from
the adjacency matrix by dividing each element $A_{ij}$ by the degree of $i$. The element $S_{ij}$ of the
stochastic matrix gives the probability that a random walker, sitting at vertex $i$, moves to $j$. The sum of the elements
of each column of $S$ is one.
Each iteration of the algorithm consists of two steps.
In the first step, called expansion, the stochastic matrix of the graph is raised to an integer power $p$
(usually $p=2$). The
entry $M_{ij}$ of the resulting matrix gives the probability that a random walker, starting
from vertex $i$, reaches $j$ in $p$ steps (diffusion flow).
The second step, which has no physical counterpart,
consists in raising each single entry of the matrix $M$ to
some power $\alpha$, where $\alpha$ is now real-valued. This operation, called inflation, enhances the weights
between pairs of vertices with large values of the diffusion flow, which are likely to be in the same community.
Next, the elements of each row must be divided by their sum, such that
the sum of the elements of the row equals one and a new stochastic matrix is recovered.
After some iterations, the process delivers a stable matrix, with some remarkable properties. Its elements are 
either zero or one, so it is a sort of adjacency matrix. Most importantly, the graph described
by the matrix is disconnected, and its connected components are the communities of the original graph.
The method is really simple to implement, which is the main reason of its success: as of now, the MCL is 
one of the most used clustering algorithms in bioinformatics. Due to the matrix multiplication of the expansion step,
the algorithm should scale as $O(n^3)$, even if the graph is sparse, as the running matrix becomes quickly 
dense after a few steps of the algorithm. However, while computing the matrix multiplication, MCL keeps only
a maximum number $k$ of non-zero elements per column, where $k$ is usually much smaller than $n$.
So, the actual worst-case running time of the algorithm is $O(nk^2)$ on a sparse graph. 
A problem of the method is the fact that the final partition is sensitive to the parameter $\alpha$
used in the inflation step. Therefore several partitions can be obtained, and it is not clear 
which are the most meaningful or representative.

\noindent {\it Maximum likelihood}. Newman and Leicht have recently proposed an algorithm based on traditional 
tools and techniques of statistical inference~\cite{newman07}. The method consists in deducing the 
group structure of the graph by checking which possible partition better ``fits'' the graph topology.
The goodness of the fit is measured by the likelihood that the observed graph structure was generated by 
the particular set of relationships between vertices that define a partition. The latter is described by two  
sets of model parameters, expressing the size of the clusters and the connection preferences among the vertices, i.e. the
probabilities that vertices of one cluster are linked to any vertex. The partition corresponding to the maximum likelihood
is obtained by iterating a set of coupled equations for the variables, starting from a suitable set of initial conditions.
Convergence is fast, so the algorithm could be applied to fairly large graphs, with up to about $10^6$ vertices.
A nice feature of the method is that it discovers more general types of vertex classes than communities. 
For instance, multipartite structure could be uncovered, or mixed patterns where 
multipartite subgraphs coexist with communities, etc..
In this respect, it is more powerful than most methods of community detection, 
which are bound to focus only on proper communities, i.e. subgraphs
with more internal than external edges. 
In addition, since partitions are defined by assigning probability values to the vertices, 
expressing the extent of their membership in a group,
it is possible that some vertices are not clearly assigned to a group, but to more groups, so the method is able to deal with 
overlapping communities. The main drawback of the algorithm is the fact that one needs to specify the number of 
groups at the beginning of the calculation, a number that is often unknown for real networks. It is possible to derive this
information self-consistently by maximizing the probability that the data are reproduced
by partitions with a given number of clusters. But this procedure involves some degree 
of approximation, and the results are often not good.

\noindent {\it L-shell method}. This is an agglomerative method designed by Bagrow and Bollt~\cite{bagrow05}.
The algorithm finds the community of any vertex, although the authors also presented a more general procedure 
to identify the full community structure of the graph.
Communities are defined locally, based on a simple criterion involving the number of edges inside and outside a group
of vertices. One starts from a vertex-origin and keeps adding vertices lying on successive shells, where a shell is defined as
a set of vertices at a fixed geodesic distance from the origin. The first shell includes 
the nearest neighbours of the origin, the second the next-to-nearest neighbours, and so on.
At each iteration, one calculates the number of edges connecting vertices of the new 
layer to vertices inside and outside the running cluster. If the ratio of these two numbers (``emerging degree'')
exceeds some predefined threshold, the vertices of the new shell are added to the cluster, otherwise the process stops.
Because of the local nature of the process, the algorithm is very fast and can identify communities
very quickly. By repeating the process starting from every vertex, one could derive a {\it membership matrix} $M$: the element
$M_{ij}$ is one if vertex $j$ belongs to the community of vertex $i$, otherwise it is zero.
The membership matrix can be rewritten by suitably permutating rows and columns based on their mutual distances.
The distance between two rows (or columns) is defined as the number of entries whose elements differ.
If the graph has a clear community structure, the membership matrix takes a block-diagonal 
form, where the blocks identify the communities. Unfortunately, the rearrangement of the matrix 
requires a time $O(n^3)$, so it is quite slow. In a different algorithm, local communities 
are discovered through greedy maximization of a local modularity measure~\cite{clauset05}. 

\noindent {\it Algorithm of Eckmann and Moses}. This is another method where communities are defined based on 
a local criterion~\cite{eckmann02}. The idea is to use the clustering coefficient~\cite{watts98} of a vertex as a 
quantity to distinguish tightly connected groups of vertices. Many edges mean many loops inside a community,
so the vertices of a community are likely to have a large clustering coefficient. The latter 
can be related to the average distance between pairs of neighbours of the vertex.
The possible values of the distance are $1$ (if neighbors are connected)
or $2$ (if they are not), so the average distance lies between $1$ and $2$. The more triangles there are in
the subgraph, the shorter the average distance. Since each vertex has always distance $1$ from its neighbours,
the fact that the average distance between its neighbours is different from $1$ reminds what happens when
one measures segments on a curved surface. Endowed with a metric, represented by the geodesic distance between
vertices/points, and a curvature, the graph can be embedded in a geometric space. Communities appear as
portions of the graph with a large curvature. The algorithm was applied to the graph representation of 
the World Wide Web, where vertices are Web pages and edges are the hyperlinks that take users from
a page to the other. The authors found that communities correspond to Web pages dealing with the same topic.

\noindent {\it Algorithm of Sales-Pardo et al.}. This is an algorithm designed to detect hierarchical
community structure (see Section~\ref{sec02}), a realistic feature of many natural, social and technological networks, that 
most algorithms usually neglect. The authors~\cite{sales07} introduce first a similarity measure between 
pairs of vertices based on Newman-Girvan modularity: basically the similarity between two vertices is the
frequency with which they coexist in the same community in partitions corresponding to 
local optima of modularity. The latter are configurations for which modularity is stable, i.e. it cannot increase
if one shifts one vertex from one cluster to another or by merging or splitting clusters.
Next, the similarity matrix is put in block-diagonal form, by minimizing a 
cost function expressing the average distance of connected vertices from the diagonal. The blocks correspond to the communities
and the recovered partition represents the largest scale organization level. To determine levels at lower scales,
one iterates the procedure for each subgraph identified at the previous level, which
is considered as an independent graph. The method yields then a hierarchy by construction, as 
communities at each level are nested within communities at higher levels. The algorithm is not fast,
as both the search of local optima for modularity and the rearrangement of the 
similarity matrix are performed with simulated annealing, but delivers
good results for computer generated networks, and meaningful partitions for some social, technological and 
biological networks.

\noindent {\it Algorithm by Rosvall and Bergstrom}. The modular structure can be considered as a reduced description
of a graph to approximate the whole information contained in its adjacency matrix.
Based on this idea, Rosvall and Bergstrom~\cite{rosvall07} envisioned
a communication process in which a partition of a network in communities represents a
synthesis $Y$ of the full structure that a signaler sends to a receiver, who tries to infer the original
graph topology $X$ from it. The best partition corresponds to the signal $Y$ that contains the most information about
$X$. This can be quantitatively assessed by the maximization of the mutual information $I(X;Y)$~\cite{shannon49}.
The method is better than modularity optimization, especially when communities are of different size.
The optimization of the mutual information is performed by simulated annealing, so the method is rather slow and can 
be applied to graphs with up to about $10^4$ vertices.

\section{Testing Methods}
\label{sec5}

When a community detection algorithm is designed, it is necessary to
test its performance, and compare it with other methods. 
Ideally, one would like to have graphs with known community structure and check whether the
algorithm is able to find it, or how closely can come to it. In any case, one needs to compare 
partitions found by the method with ``real'' partitions.
How can different partitions of the same graph be compared?
Danon et al.~\cite{danon05}
used a measure borrowed from information theory, the {\it normalized mutual information}.
One builds a {\it confusion matrix} $N$, whose element $N_{ij}$ is the number of vertices of the real community 
$i$ that are also in the detected community $j$. Since the partitions to be compared may have different numbers of 
clusters, $N$ is usually not a square matrix. The similarity of two partitions $A$ and $B$ is given by the following expression
\begin{equation}
I(A,B)=\frac{-2\sum_{i=1}^{c_A}\sum_{j=1}^{c_B}N_{ij}\log(N_{ij}N/N_{i.}N_{.j})}
{\sum_{i=1}^{c_A}N_{i.}\log(N_{i.}/N)+\sum_{j=1}^{c_B}N_{.j}\log(N_{.j}/N)},
\label{eqt0}
\end{equation}
where $c_B$ ($c_A$) is the number of communities in partition $A$ ($B$), $N_{i.}$ is the sum of the elements 
of $N$ on row $i$ and $N_{.j}$ is the sum of the elements of $N$ on column $j$. 
Another useful measure of similarity between partitions is the {\it Jaccard index}, which is 
regularly used in scientometric research. Given two partitions $A$ and $B$, the Jaccard index is defined as
\begin{equation}
I_J(A,B)=\frac{n_{11}}{n_{11}+n_{01}+n_{10}},
\label{eqt1}
\end{equation}
where $n_{11}$ is the number of pairs of vertices which are in the same community in both partitions and $n_{01}$
($n_{10}$) denotes the number of pairs of elements which are put in the same community in $A$ ($B$) and in
different communities in $B$ ($A$).
A nice presentation of 
criteria to compare partitions can be found in Ref.~\cite{gustafsson06}.

\begin{figure}
\begin{center}
\includegraphics[width=9cm]{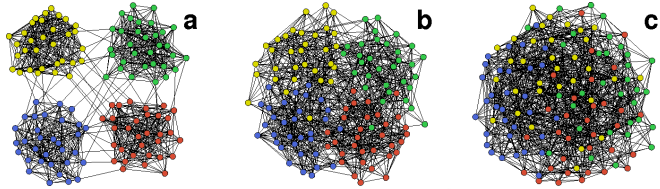}
\caption {\label{Figure11} Benchmark of Girvan and Newman. The three pictures correspond 
to $z_{in}=15$ (a), $z_{in}=11$ (b) and $z_{in}=8$ (c). In (c) the four groups are basically 
invisible. Reprinted figure with permission from Guimer\`a R, Amaral LAN, Nature 433, 895 (2005). Copyright 2005 by the
Nature Publishing Group.}
\end{center}
\end{figure}

In the literature on community detection, algorithms have been generally 
tested on two types of graphs: computer generated graphs and real networks.
The most famous computer generated benchmark is a class of graphs designed by Girvan and Newman~\cite{Girvan:2002}.
Each graph consists of $128$ vertices, arranged in four groups with $32$ vertices each: $1-32$, $33-64$, $65-96$ and $97-128$. 
The average degree of each vertex is set to $16$.
The density of edges inside the groups is tuned by a parameter $z_{in}$, expressing the average number of 
edges shared by each vertex of a group with the other members (internal degree). Naturally, when $z_{in}$ is close to $16$,
there is a clear community structure (see Fig.~\ref{Figure11}a), as most edges will join vertices of the same community, 
whereas when $z_{in}<=8$ there are more
edges connecting vertices of different communities and the graph looks fuzzy (see Fig.~\ref{Figure11}c). 
In this way, one can realize different degrees of mixing between the groups. 
In this case the test consists in calculating the similarity between the partitions determined by the
method at study and the natural partition of the graph in the four equal-sized groups. The similarity can be calculated
by using the measure of Eq.~\ref{eqt0}, but in the literature one used a different quantity, i.e. the fraction of correctly 
classified vertices. A vertex is correctly classified if it is in the same cluster with at least $16$ of its ``natural'' partners.
If the model partition has clusters given by the merging of two or more natural groups, all vertices of the cluster
are considered incorrectly classified. The number of correctly classified vertices is then divided by the total size of the 
graph, to yield a number between $0$ and $1$. One usually builds many realizations of the graph for a particular value of 
$z_{in}$ and computes the average fraction of correctly classified vertices, which is a measure of the 
sensitivity of the method. The procedure is then iterated for different values of $z_{in}$. Many different 
algorithms have been compared with each other according to the diagram where
the fraction of correctly classified vertices is plotted against $z_{in}$. Most algorithms usually do a good job 
for large $z_{in}$ and start to fail when $z_{in}$ approaches $8$.
The recipe to label vertices as 
correctly or incorrectly classified is somewhat arbitrary, though, and measures like those of 
Eqs.~\ref{eqt0} and~\ref{eqt1} are probably more objective.
There is also a subtle problem concerning the reliability of the test. Because of the randomness involved in the
process of distributing edges among the vertices, it may well be that, in specific realizations of the graph, 
some vertices share more edges with members of another group than of their own. In this case, it is inappropriate to 
consider the initial partition in four groups as the real partition of the graph.

Tests on real networks usually focus on a limited number of examples,
for which one has precise information about the vertices and their properties. 

\begin{figure}
\begin{center}
\includegraphics[width=7cm]{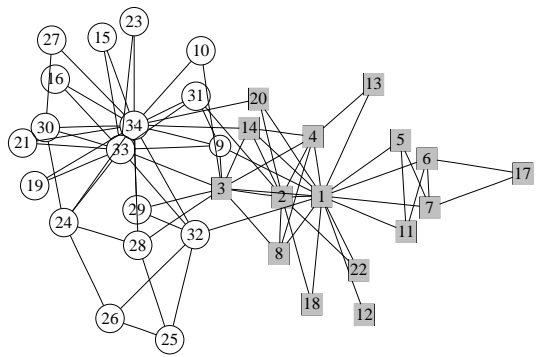}
\caption {\label{Figure12}Zachary's karate club network, an example of graph with known community structure.  
Reprinted figure with permission 
from Newman MEJ, Girvan M, Physical Review E 69, 026113, 2004. Copyright 2004 by the Americal Physical Society.}
\end{center}
\end{figure}

The most popular real network with a known community structure is 
the social network of Zachary's karate club (see Fig.~\ref{Figure12}).
This is a social network representing the personal relationships between members of a karate club at an American 
university. During two years, the sociologist Wayne Zachary observed the ties between members, both inside and outside 
the club~\cite{zachary}. At some point, a conflict arose between the club's administrator (vertex $1$) 
and one of the teachers (vertex $33$), which 
led to the split of the club in two smaller clubs, with some members staying with the administrator and the others
following the instructor. Vertices of the two groups are highlighted by squares and circles 
in Fig.~\ref{Figure12}. The question is whether the actual social split could be predicted from the network topology.
Several algorithms are actually able to identify the two classes, modulo a few intermediate vertices, which may be misclassified
(e.g. vertices $3$, $10$). Other methods are less successful: for instance, the maximum of Newman-Girvan modularity corresponds
to a split of the network in four groups~\cite{Duch:2005,donetti04}.
It is fundamental however to stress that the comparison of community
structures detected by the various methods with the split of Zachary's
karate club is based on a very strong assumption: that the split actually
reproduced the separation of the social network in two communities.
There is no real argument, beyond common wisdom, supporting this assumption.

Two other networks have frequently been used to test community detection algorithms: the network of American college football teams
derived by Girvan and Newman~\cite{Girvan:2002} and the social network
of bottlenose dolphins constructed by Lusseau~\cite{lusseau:2003}.
Also for these networks the caveat applies: Nothing guarantees that
``reasonable'' communities, defined on the basis of non-topological information,
must coincide with those detected by methods based only on topology.

\section{The Mesoscopic Description of a Graph}
\label{sec6}

Community detection algorithms have been applied to a huge variety of real systems, including social, biological and
technological networks. The partitions found for each system are usually similar, as the
algorithms, in spite of their specific implementations, 
are all inspired by close intuitive notions of community.
What are the general properties of these partitions? The analysis of partitions and their properties
delivers a {\it mesoscopic description} of the graph, where the communities, and not the vertices, are the elementary 
units of the topology. The term mesoscopic is used because the relevant scale here lies between the scale of the vertices and
that of the full graph.
A simple question is whether the communities 
of a graph are usually about of the same size or whether the community sizes have some special distribution.
It turns out that the 
distribution of community sizes is skewed, with a tail 
that obeys a power law with exponents in the range between 
$1$ and $3$~\cite{Palla:2005,newman04rev, arenasrev, Clauset:2004}. So, there seems to be no characteristic
size for a community: small communities usually coexist with large ones.
\begin{figure}
\begin{center}
\includegraphics[width=8cm]{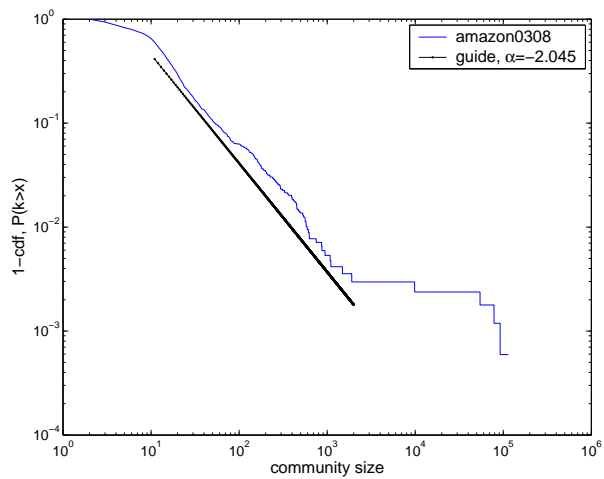}
\caption {\label{Figure13} Cumulative distribution of community sizes for the 
Amazon purchasing network. The partition is derived by greedy modularity optimization.  
Reprinted figure with permission 
from Clauset A, Newman MEJ and Moore C, Physical Review E 70, 066111, 2004. Copyright 2004 by the Americal Physical Society.}
\end{center}
\end{figure}
As an example, Fig.~\ref{Figure13} shows the cumulative distribution of community sizes for 
a recommendation network of the online vendor Amazon.com. Vertices are products and 
there is a connection between item $A$ and $B$ is $B$ was frequently purchased by buyers of $A$.
We remind that the cumulative distribution is the integral of the probability distribution: if the 
cumulative distribution is a power law with exponent $\alpha$, the probability distribution is also 
a power law with exponent $\alpha+1$. 

If communities are overlapping, one could derive a network, where the
communities are the vertices and pairs of vertices are connected if their corresponding communities
overlap~\cite{Palla:2005}. Such networks seem to have some special properties. For instance, 
the degree distribution is a particular function, with an initial exponential
decay followed by a slower power law decay. A recent analysis has shown that such
distribution can be reproduced by assuming that the graph grows 
according to a simple preferential attachment mechanism, where communities
with large degree have an enhanced chance to interact/overlap with new communities~\cite{pollner}.

Finally, by knowing the community structure of a graph, it is possible to classify vertices according
to their roles within their community, which may allow to infer individual properties of the
vertices. A nice classification has been proposed by Guimer\'a and 
Amaral~\cite{Guimera:2005,Guimera:2005b}. The role of a vertex depends on the values of 
two indices, the $z$-score and the participation ratio, that determine the position of the vertex
within its own module and with respect to the other modules. The $z$-score compares the internal degree
of the vertex in its module with the average internal degree of the vertices in the module.
The participation ratio says how the edges of the vertex are distributed among the modules.
Based on these two indices, Guimer\'a and Amaral distinguish seven roles for a vertex.
These roles seem to be correlated to functions of vertices: in metabolic networks, for instance,
vertices sharing many edges with vertices of other modules (``connectors'') are often
metabolites which are more conserved across species than other metabolites,  
i.e. they have an evolutionary advantage~\cite{Guimera:2005}.  

\section{Future Directions}
\label{sec7}

The problem of community detection is truly interdisciplinary. It involves scientists
of different disciplines both in the design of algorithms and in their applications.
The past years have witnessed huge progresses and novelties in this topic. Many methods 
have been developed, based on various principles. 
Their scalability has improved by at least one power in the 
graph size in just a couple of years.
Currently partitions in graphs  with up to millions of vertices can be found.
From this point of view, the limit is close, and 
future improvements in this sense are unlikely. Algorithms running in linear time are very quick,
but their results are often not very good. 

The major breakthrough introduced by the new methods is the possibility of extracting 
graph partitions with no preliminary knowledge or inputs about the community
structure of the graph. Most new algorithms
do not need to know how many communities there are, a major drawback of computer science approaches:
they derive this information from the graph topology itself. Similarly, algorithms of new generation
are able to select one or a few meaningful partitions, whereas social science approaches 
usually produce a whole hierarchy of partitions, which they are unable to discriminate.
Especially in the last two years, the quality of the output produced by some algorithms
has considerably improved. Realistic aspects of community structure, like overlapping 
and hierarchical communities, are now often taken into account.  

The main question is: is there at present a good method to detect communities in graphs?
The answer depends on what is meant by ``good''. Several algorithms give 
satisfactory results when they are tested as described in Section~\ref{sec5}: in this respect,
they can be considered good. However, if examined in more detail, some methods disclose 
serious limits and biases.
For instance, the most popular method used nowadays, modularity optimization,
is likely to give problems in the analysis of large graphs. 
Most algorithms are likely to fail in some limit, still one can derive useful
indications from them:
from the comparison of partitions derived by different methods one could extract the cores of real communities.
The ideal method is one that delivers meaningful partitions and handles overlapping communities and 
hierarchy, possibly in a short time. No such method exists yet.  

Finding a good method for community detection 
is a crucial endeavour in biology, sociology and computer science.
In particular, biologists often rely on the application of clustering techniques to classify their data. Due to 
the bioinformatics revolution,
gene regulatory networks, protein-protein interaction networks, metabolic networks, etc., are 
now much better known that they used to be in the past and finally susceptible to 
solid quantitative investigations. Uncovering their modular structure 
is an open challenge and a necessary step to discover properties of elementary biological constituents
and to understand how biological systems work.

\end{document}